\documentclass[twocolumn,showpacs,preprintnumbers,longbibliography]{revtex4-2}
\usepackage{amsfonts}
\usepackage{amssymb}
\usepackage{amsmath}
\usepackage{graphicx}
\usepackage{dcolumn}
\usepackage{bm}
\usepackage{color}
\usepackage{braket}
\usepackage[colorlinks=true, allcolors=blue]{hyperref}
\usepackage{float}
\usepackage{appendix}

\usepackage[utf8]{inputenc}

\begin{document}
\title{Photon Squeezing and Its Signatures of Quantum Phase Transitions in the Open Quantum Rabi-Stark Model}
\author{Tian Ye$^{1}$ }
\author{Xinghan Chen$^{2}$}
\author{Chen Wang$^{2,}$}
\email{wangchen@zjnu.cn}	
\affiliation{
$^1$ Anhui Province Key Laboratory for Control and Applications of Optoelectronic Information Materials,  Department of Physics, Anhui Normal University, Wuhu 241000, China \\
$^2$ Department of Physics, Zhejiang Normal University, Jinhua 321004, China
}

\date{\today }

\begin{abstract}
As a hallmark of nonclassical light, squeezed light is of profound theoretical interest and holds broad practical promise for emerging quantum technologies. In this work, we investigate steady-state optical quadrature squeezing in the open quantum Rabi-Stark model by employing the quantum dressed master equation. Both numerically and analytically, we find that positive (negative) Stark coupling tends to enhance (suppress) the squeezing effect. The quadrature squeezing exhibits distinct signatures associated with both first- and second-order quantum phase transitions (QPTs). 
Notably, a sharp vanishing of squeezing is observed across the first-order QPT, suggesting its potential as a sensitive probe of such transitions. In the vicinity of the second-order QPT, we further demonstrate that the squeezing factor displays finite-size scaling behavior, indicating a promising route toward the realization of near-perfect squeezing. Moreover, we establish a quantitative criterion for the disruption of quantum criticality induced by thermal fluctuations, which may offer valuable guidance for future experiments. These findings contribute to a deep understanding of nonclassical light in light-matter interacting systems and provide useful insights for the design of strong optical squeezing states.
\end{abstract}

\maketitle

\section{\label{Introduction} Introduction }
The light-matter interacting systems are widely regarded as one kind of representative models in quantum optics and quantum information, and have been attracting extensive attention for decades~\cite{scully1997book,haroche2006book,diaz2019rmp,nori2019nrp}. Due to the tremendous progresses of its simulation platform with solid-state settings such as the superconducting qubits ~\cite{yoshihara2017np,blais2021rmp}, trapped ions~\cite{trapped_ion2018prx,duan2021nc,xyzhao2025prl}, and the cold atoms~\cite{coldatom2017nc,koch2023nc}, the qubit-photon systems reach the uncharted regimes of ultrastrong and even deepstrong couplings.
In the ultrastrong coupling regime, the celebrated rotating-wave approximation becomes invalid, and the qubit-photon coupled systems then exhibit a distinguishing physics, characterized by hybridizing qubit and photon. 
It spurs plenty of influential work on ultrastrong coupled qubit-photon systems, ranging from the exact and approximated analytical solutions\cite{BSshift2010prl,braak2011prl,chen2012pra,braak2016jpa,leicong2017pra}, multiphoton Rabi oscillation dynamics\cite{garziano2015pra,garziano2016prl}, nonclassical states and photon blockade\cite{ridolfo_photonblockade,qbin2021prl,yxzhang2025adv_qm,yxzhang2026oe,rhzheng2023prl}, quantum phase transition (QPT)~\cite{hwang2015prl,mxliu2017prl,xychen2021pra,mxliu2023pra}, to quantum metrology\cite{garbe2020prl}.  

The quantum Rabi model (QRM)~\cite{braak2011prl,rabi_rabimodel},  serving as a paradigm for quantum light-matter interactions, describes the simplest qubit-photon coupling system, which is composed of a single-mode photon field coupled to a qubit (a two-level quantum system).
In spite of its few components, the QRM can exhibit a QPT in the effective thermodynamical limit as the frequency ratio of the qubit and photon field tends to infinity~\cite{hwang2015prl}. 
Such a QPT is characterized by the macroscopic excitation of the photon field, and is usually named as superradiant phase transition (SRPT), retaining the name of the Dicke model counterpart. 
The non-classical light can be generated not only via traditional nonlinear optical processes, e.g., parametric down conversion, but also by coupling a conventional light field to a nonlinear quantum system, such as a qubit. 
As unveiled in the QRM, the coupling of quantum light to a qubit may reshape the harmonic-oscillator-type light field to plenty of nonclassical states, including photon blockade and photon quadrature squeezing~\cite{hwang2010pra,mxliu2020adv_qm,ridolfo_photonblockade,qbin2021prl,xdh2026pra}. 
In particular, 
as the QRM approaches to the critical point, the position-quadrature fluctuation diverges along with the macroscopic excitation of the superradiant phase transition.
Meanwhile, the momentum-quadrature fluctuation vanishes for the minimum-uncertainty-state nature of the critical photon field~\cite{hwang2015prl}.  
Hence, the critical photon field of the QRM is then promising for strong squeezing state preparation. 

Practically, the influence of thermal noise during the measurement on nonclassicality of photon is inevitable.
Accordingly, the influence of thermal fluctuation should be included. The framework of quantum dressed master equation (DME)~\cite{blais_dME} will generally be applied to circumvent the limitation of the standard optical master equation with bare jump operator. Specifically, the standard optical master equation characterizes dissipation through quantum jumps between eigenstates of the uncoupled system components and further results in unphysical predictions, such as zero-temperature-environment-induced excitation out of the ground state, particularly in ultrastrong coupling regimes~\cite{blais_dME,diaz2019rmp,nori2019nrp,boite2020aqt}. 

When simulating the QRM in cavity QED platform~\cite{Parkins2013pra,Parkins2014pra} or trapped ion set-up~\cite{lcong2023pra}, a nonlinear term, Stark term, may appear. 
The Stark term can be tuned independently based on the above schemes, which gives rise to a new generation of the QRM, i.e., the quantum Rabi Stark model (QRSM). 
Given the extensive demonstrations of optical squeezing via nonlinear interactions~\cite{scully1997book,squeeze_exp1,squeeze_exp2}, the QRSM provides a promising route to strengthen the photonic quadrature squeezing comparing with the scenario of the standard QRM. 
Furthermore, the QRSM has been shown to exhibit a variety of singular behaviors, including both the first- and second-order QPTs, and spectral collapse~\cite{yfxie2019jpa,yfxie2020pra,xychen2020pra}. 
These singular behaviors typically reshape the quantum characteristics of the system, among which quantum correlation and entanglement have been widely investigated.
However, photonic nonclassicality has attracted far less attention. Hence, it is highly desirable to explore the potential photonic-nonclassicality signature associated with these singular behaviors under the interplay of strong light-matter interactions and quantum dissipation.

In this study, using the DME, we find not only enhanced squeezing induced by a positive Stark term but also optical-squeezing signatures associated with both first- and second-order QPTs. Furthermore, the steady-state optical squeezing in the critical regime exhibits scalability with the effective system size, suggesting the potential for near-perfect squeezing.
The remainder of this paper is structured as follows. Section~\ref{sec2} briefly introduces the QRSM, the DME, and the definition of optical quadrature squeezing. In Sec.~\ref{sec3}, we analytically unveil the influence of the Stark term and the first-order QPT on optical squeezing. Section~\ref{sec4} demonstrates the criticality-enhanced squeezing and its suppression under thermal noise. Finally, Sec.~\ref{conclusion} presents our conclusions. In addition, Appendix~\ref{appendix1} provides an analytical derivation of the quadrature squeezing for the QRSM ground state, while Appendix~\ref{appendix2} analytically examines the critical behavior of this squeezing in the effective thermodynamic limit.


\section{\label{sec2}model and method}

\subsection{\label{model}Quantum Rabi-Stark model}
The QRSM generalizes the QRM by including an extra nonlinear qubit–cavity coupling, with Hamiltonian~\cite{Parkins2013pra,EckleRS2017}
\begin{equation}\label{HRS}
H_{\text{RS}}\ =\omega _{0}a^{\dagger }a+\frac{\Delta }{2}\sigma
_{z}+Ua^{\dagger }a\sigma _{z}+g\sigma _{x}(a+a^{\dagger }),
\end{equation}%
where  $a$ ($a^{\dagger}$) is the creation (annihilation) operator of the cavity field, $\sigma_{x,y,z}$ the Pauli matrices of the two-level atom (qubit),
and $U$ is the strength of the nonlinear coupling. The cavity frequency $\omega_{0}$ is set to unity for simplicity, while $\Delta$ and g denote the qubit splitting and the linear qubit-cavity coupling strength, respectively.

Notably, although the term $a^{\dagger}a\sigma_{z}$ can arise when approximating the QRM to the Bloch–Siegert Hamiltonian in the intermediate-coupling regime with $g\ll\left(\omega_0+\Delta\right)$ ~\cite{BSshift2010prl,blais_dME,zueco2009pra}, it is treated as an independent tunable parameter in the QRSM. 
The Stark term can be simulated by the resonant Raman transition in $^{87}$Rb atom interacting with high finesse cavity mode~\cite{Parkins2013pra}. Furthermore, the Stark coupling strength is constrained to $\vert U\vert\leq 1$, since  $ U>1$ ($U<-1$) would lead to an unphysical negative frequency for the polariton associated with the ground-state (or excited-state) qubit, resulting in an unbounded ground-state energy.

Moreover, the QRSM presents multiple singular behaviors, including both the first- and second-order QPTs and the spectral collapse. Specifically, in the QRSM with positive Stark coupling, a first-order QPT can occur at the qubit–cavity coupling strength $g_{c}^{1}=\sqrt{{(1-U^{2})\Delta }/{2U}}$~\cite{yfxie2019jpa,yfxie2020pra}. 
Furthermore, in the effective thermodynamical limit of $U \rightarrow \pm 1$,
not only do the energy spectra tend to collapse to $E_c^\pm=\mp\Delta/2-2g^2$ with at most some discrete energy levels lying below this energy~\cite{yfxie2019jpa}; but also the ground state would present a second-order SRPT when the dipole coupling exceeds the critical one $g_c^\pm=\sqrt{1\mp\Delta/2}$~\cite{yfxie2019jpa,xychen2021pra}.

\subsection{\label{ME} Quantum dissipation and steady state}
In fact, the quantum dissipation induced environment is inevitable for the quantum system. To describe the effect, the
qubit and the photon field in the QRSM are coupled individually to two bosonic thermal baths. The total system-environment Hamiltonian can be denoted as
\begin{equation}
H_{\text{total}}\ =H_{\text{QRSM}}+H_\text{B}+V,
\end{equation}%
where $H_\text{B}$ is the Hamiltonian for thermal baths with $
H_\text{B}=\sum _{u=\text{at},\text{c};\,k}\omega_{u,k}b_{u,k}^{\dagger }b_{u,k}$
and $b_{u,k}^{\dagger}$ ($b_{u,k}$) the annihilation (creation) operator of $\omega
_{u,k}$-frequency boson mode in the $u$th bath. Besides, $V$ is the system-environment
interaction Hamiltonian. Its component $V_\text{at}$ ($V_\text{c}$)
describes the interaction between qubit (photon field) and bosonic thermal bath respectively, i.e.,
\begin{subequations}
\begin{align}
V_\text{q}=& \sum _{k}\lambda _{\text{q},k}(b_{\text{q},k}+b_{\text{q},k}^{\dagger })\sigma _{x},~
\label{vat} \\
V_\text{c}=& \sum _{k}\lambda _{\text{c},k}(b_{\text{c},k}+b_{\text{c},k}^{\dagger })(a+a^{\dagger
}),~  \label{vc}
\end{align}%
\end{subequations}
with $\lambda_{\text{q},k}~(\lambda_{\text{c},k})$ the strength of the interaction between the 
qubit (photon field) and the corresponding bosonic bath. Then, the spectral function
of system-environment interaction defined as %
$\gamma_{\text{q}(\text{c})}(\omega)=2\pi \sum_{k}|\lambda_{\text{q}(\text{c}),k}|^{2}\delta_{\omega,\omega_{k}}$%
is taken as the Ohmic spectrum case~\cite{uweiss2012book}, 
i.e., $\gamma_{\text{q}}(\omega)=\alpha_{\text{q}}{\omega e^{-{\omega}/{\omega _{c}}}/\Delta} $ and 
$\gamma_{\text{c}}(\omega )=\alpha_{\text{c}}{\omega }e^{-{\omega }/{\omega _{c}}/{\omega_{0}}}$ %
with $\alpha_{\text{q}(\text{c})}$ the system-environment interaction strength and
$\omega_{c}$ is the cutoff frequency of two thermal baths.

To avoid the unphysical predictions of the standard optical master equation as mentioned in the introduction, 
The dissipative dynamics of the QRSM should be investigated in the basis of eigenstates of the QRSM (the dressed-state picture). The system-environment interactions are then expanded in the dressed-state picture as
\begin{subequations}
\begin{align}
V_{\text{q}}&=\sum_{k,m,n}\lambda _{\text{q},k}(b_{\text{q},k}+b_{\text{q},k}^{\dagger })L_{mn}^\text{q} \\
V_\text{c}&=\sum _{k,m,n}\lambda _{\text{c},k}(b_{c,k}+b_{c,k}^{\dagger })L_{mn}^\text{c}
\end{align}
\end{subequations}
with the jump operators $L^\text{q}_{mn}={\langle}\phi_m|\sigma_x|\phi_n{\rangle}|\phi_{m}\rangle\langle\phi_{n}|$,
$L^\text{c}_{mn}={\langle}\phi_m|(a^\dag+a)|\phi_n{\rangle}|\phi_{m}\rangle\langle\phi_{n}|$,
and $|\phi_{m(n)}\rangle$ the eigenstate of the QRSM. 
Finally, under the assumption of weak system-environment interaction and short correlation time of the environment, we can
apply the celebrated Born-Markov approximation and obtain the DME~\cite{blais_dME,boite2020aqt,nori2019nrp}, 
\begin{eqnarray}
~\frac{\partial\rho_s}{\partial t}&&=-i\left[H_\text{RS},\rho_{s} \right] \notag \\%
&&+\sum_{m,n>m}\{\Gamma_{u}^{n,m}\left[ 1+n_u\left(\Delta_{n,m}\right) \right]%
 \mathcal{D}[\phi_{m}\rangle \langle \phi_{n}|,\rho_{s}]  \notag \\%
&&+\Gamma_{u}^{m,n}n_{u}\left( \Delta_{n,m}\right) \mathcal{D}[|\phi_{n}\rangle \langle %
\phi_{m}|,\rho_{s}]\},
\end{eqnarray}%
where the Lindblad dissipator is $\mathcal{D}[O,\rho_{s}]=\frac{1}{2}(2O\rho_{s}O^{\dagger }%
-O^{\dagger }O\rho_{s}-\rho_{s}O^{\dagger }O)$ 
with $\rho_{s}$ the reduced density matrix of the system and $|\phi_{m(n)}\rangle $ the eigenstate with eigenenergy ${E}_{m(n)}$ of the QRSM, $\Delta_{n,m}=E_{n}-E_{m}$ is the energy gap between different eigenstates, $n_{u}\left(\Delta_{n,m}\right)=1/(e^{\Delta_{n,m}/k_\text{B}T_u}-1)$ denotes the Bose-Einstein distribution function, and the effective transition rates $\Gamma_\text{q}^{m,n}$ and $\Gamma_\text{c}^{m,n}$ are given by
\begin{subequations}
\begin{align}
\Gamma_\text{q}^{m,n}=& \alpha _{q}\frac{\Delta_{n,m}}{\Delta }%
e^{-\frac{\Delta_{n,m}}{\omega _{c}}}|\langle \phi_{n}
|(\sigma _{-}+\sigma _{+})|\phi_{m}\rangle |^{2}, \\
\Gamma_\text{c}^{m,n}=& \alpha _{c}\frac{\Delta_{n,m}}{\omega _{0}%
}e^{-\frac{\Delta_{n,m}}{\omega _{c}}}|\langle\phi_{n}
|(a+a^{\dagger })|\phi_{m}\rangle |^{2}.
\end{align}%
\end{subequations}
The DME performs particularly well for long-time dissipative dynamics even at ultra-strong and deep-strong qubit-photon couplings~\cite{Ridolfo2018pra,boite2020aqt,nori2019nrp}, enabling our study of steady-state photon squeezing based on the DME.

\subsection{\label{squeezing}Quadrature squeezing of photons}

In quantum optics, the nonclassical characterization of the optical field can be represented by a pair of canonically conjugate field quadratures, i.e.,
~$X_\theta=ae^{-i\theta}+a^\dagger e^{i\theta}$ and $P_\theta=-i\left(ae^{-i\theta}-a^\dagger e^{i\theta}\right)$.  Then, the inherent quantum noise of photon field is bounded from below by the celebrated Heisenberg uncertainty principle, i.e.,~$\Delta X_\theta\Delta P_\theta\geq 1$ with $\Delta \mathcal{O}=\sqrt{\langle \mathcal{O}^2\rangle-\langle \mathcal{O}\rangle^2}$, with the optical operator $\mathcal{O}$. Considering the ubiquitous coherent states, they are easy to be found a kind of minimum uncertainty states with $\Delta X_\theta=\Delta P_\theta=1$. 
Accordingly, $\Delta X_\theta=1$ is regarded as a lower limit of quantum noise prior to the proposal of squeezing states. 
Historically, the first type of squeezed states, the squeezed coherence state is found by performing the squeezing transformation $S(\xi)=e^{(\xi^*a^2-\xi {a^\dagger}^2)/2}$ on the coherent state $|\alpha\rangle$~\cite{squeeze_orgn1,squeeze_orgn2,squeeze_orgn3}. The formal theoretical proposal of squeezed states came a few years later, along with the potential application of improving the sensitivity of interferometers to breaking through the standard quantum limit~\cite{squeeze_orgn4}.  
The most common squeezing state is the squeezing vacuum state which can be obtained by setting $\alpha=0$ in the squeezing coherence state.  

Finally, the degree of quadrature squeezing of photons can be quantified by the principal quadrature squeezing factor
\begin{equation}
\xi_\text{B}^2=\min_{\theta\in[0,2\pi)}\{ (\Delta X_\theta)^2\}. \label{xi_def}
\end{equation}%
Clearly, $\xi_\text{B}^2=1$ for the coherent state, and $\xi_\text{B}^2<1$ implies the squeezing of photons. Moreover, the representative squeezing states and squeezing vacuum states exhibit $\xi_\text{B}^2=e^{-|r|}<1$, consistent with the condition $\xi_\text{B}^2<1$. 

In the experiment, the quadrature squeezing of photons is typically quantified using the balanced homodyne detection~\cite{scully1997book}.
In this scheme, a 50:50 beam splitter mixes the signal with a strong reference laser onto two identical photodetectors, and their output currents are subtracted to yield the current directly proportional to the signal-field quadrature component selected by the reference-laser phase.  Crucially, the balanced homodyne detection eliminates common-mode noise of the reference laser, enabling the precise quantification of quantum fluctuation in the selected field quadrature~\cite{squeeze_detect}.
\begin{figure}[tbp]
\centering
\includegraphics[width=\linewidth]{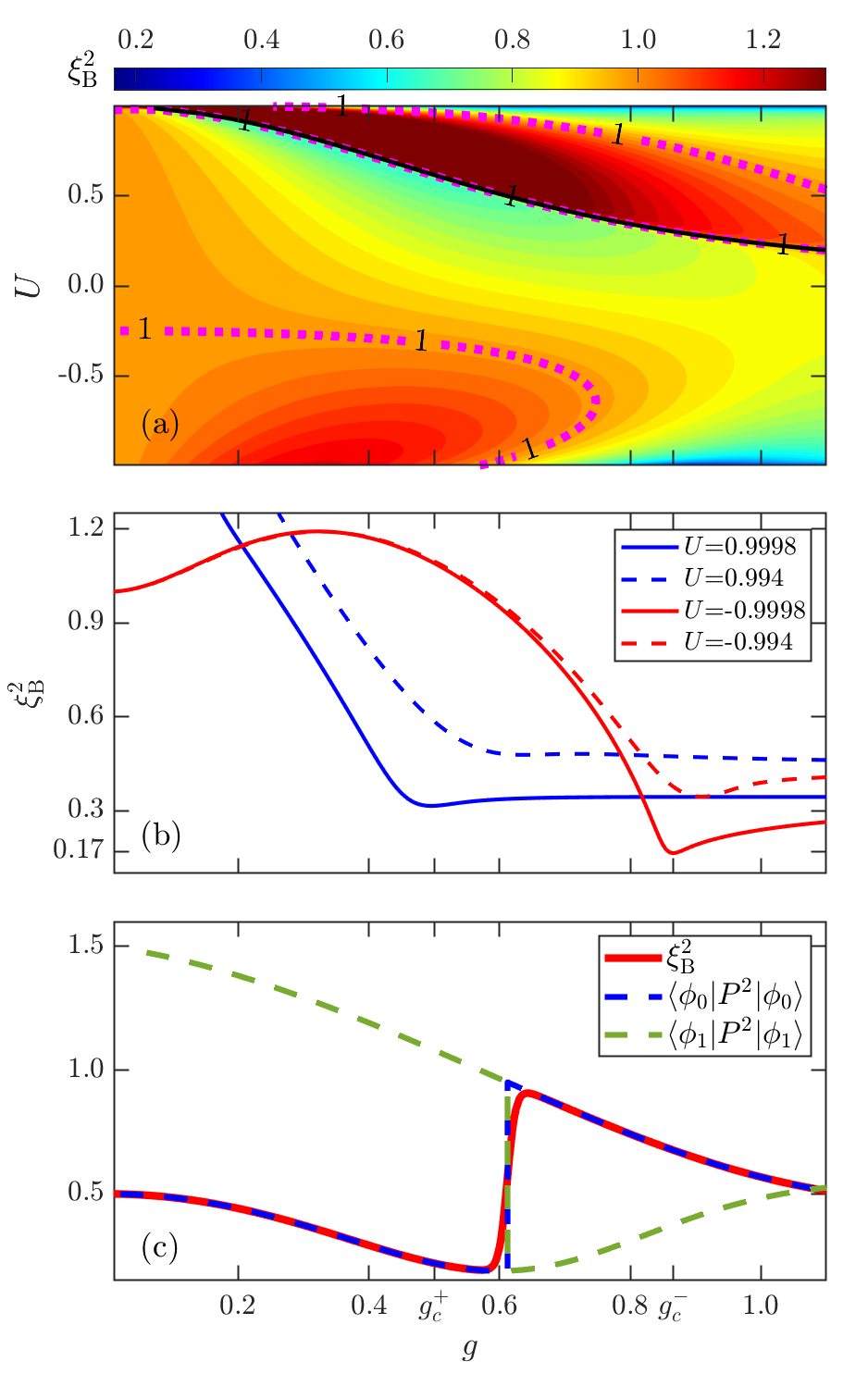}
\caption{The quadrature squeezing factor $\xi_\text{B}^2$ of the steady state in the dissipative QRSM.
(a) $\xi_\text{B}^2$ as a function of dipole-coupling strength $g$ and Stark-coupling strength $U$. The magnetic dash line denotes the contour of $\xi_\text{B}^2=1$, which is the boundary the squeezing photon field with $\xi_\text{B}^2<1$ and the normal photon field with $\xi_\text{B}^2<1$. The black solid line denotes the position of first-order QPT in the closed QRSM. (b) $\xi_\text{B}^2$ with 
near-unit $|U|$ extracted from the top and bottom of panel (a). (c) $\xi_\text{B}^2$ across the first-order QPT with $U=0.5$, along with the expectation of momentum-quadrature square $P^2$ at the ground state $\phi_{0}$ and the first-excited state $\phi_{1}$. Besides, the other parameters are given by $\omega_0=1$, $\Delta=0.5$, $\alpha_\text{q}=\alpha_\text{c}=10^{-4}$, $\omega_{c}=10$, and $k_\text{B}T_\text{q}=k_\text{B}T_\text{c}=0.003$.}
\label{fig1}
\end{figure}

\section{\label{sec3}photon squeezing induced by nonlinear Stark coupling}
We investigate the steady-state optical squeezing of the dissipative QRSM,
where the squeezing of photons is quantified with the quadrature squeezing factor
$\xi_\text{B}^2$ as defined in Eq.~(\ref{xi_def}). Fig.~\ref{fig1}(a) shows
$\xi_\text{B}^2$
as a function of the linear qubit-photon coupling strength $g$
and Stark-coupling strength $U$.
To distinguish the regions of photon squeezing, where with $\xi_\text{B}^2<1$, 
from those of a normal photonic field, where   $\xi_\text{B}^2>1$,
we plot the contour of $\xi_\text{B}^2=1$ using the magnetic dashed lines.
Obviously, a wide parameter region excluding the red area enclosed by the magenta dashed contour exhibits  photon squeezing. 

In the dissipative standard QRM, corresponding to $U=0$ in Fig.~\ref{fig1}(a), the photon squeezing is quite weak, with $\xi_\text{B}^2\gtrsim 0.9$. However,
when a strong Stark coupling with near-unit $\vert U\vert$ is present, the photon squeezing can be strengthened significantly, reaching $\xi_\text{B}^2{\approx}0.17$,
if the qubit-cavity coupling  $g$ is large enough.
Such behaviors are illustrated by the narrow blue regions
in the top-left and bottom-left corners of Fig.~\ref{fig1}(a) and Fig.~\ref{fig1}(b).

We  defer the detailed discussion of this strong photon squeezing at $|U|\approx 1$ to the next section, as it involves the critical behaviors of the second-order SRPT.
It is worth noting that a positive Stark coupling can also enhance the squeezing of photons
to $\xi_\text{B}^2\approx 0.7$, as shown by the cyan region on the left of the black solid line of Fig.~\ref{fig1}(a). In contrast,  a negative Stark coupling tends to suppress  photon squeezing over a broad parameter region.

It is quite interesting to find that the photon squeezing in the cyan region  vanishes sharply when $g$ exceeds a critical value, which is highly consistent with the first-order QPT position in the closed QRSM, as also denoted by the black solid line in Fig.~\ref{fig1}(a). 
To confirm this, we examine $\xi_\text{B}^2$ across the first-order QPT with $U=0.5$, as shown in Fig.~\ref{fig1}(c). Meanwhile, the sharp transition of $\langle \psi_0 |P^2| \psi_0\rangle$ and $\langle \psi_1 |P^2| \psi_1\rangle$ implies the energy-level crossing between $|\psi_0\rangle$ and $|\psi_1\rangle$, i.e., the first-order QPT. The squeezing factor $\xi_\mathrm{B}^2$ at low-temperature steady state decreases along the momentum quadrature as $g$ increases. However, 
as $g$ crosses the critical point, $\xi_\mathrm{B}^2$ rises dramatically. 

Experimentally, both the fine-tuning to a first-order QPT point and the resolution of the near-vanishing energy gap around the point are extremely demanding, which hinders the identification of the transition via direct energy-gap measurements. Fortunately, the step-like change between the squeezing state and the normal state across the QPT could be measured by the balanced homodyne detection, as mentioned in Sec.~\ref{squeezing}, and then the squeezing signal above may be helpful for probing the first-order QPT. Therefore, the Stark coupling surely enriches photon squeezing behaviors.

To elucidate the underlying mechanism, 
we first include the analytical
expression~\cite{jianma2011physrep}
\begin{equation}\label{xi_min1}
\xi_\text{B}^2=1+2\left(\langle a^\dagger a\rangle-\vert\langle a\rangle\vert^2\right)-2\vert\langle a^2\rangle-\langle a\rangle^2\vert.
\end{equation}
Considering the parity properties of the QRSM and non-negative behavior of $\langle a^2\rangle$ at the steady state, we simplify the expression of $\xi_\text{B}^2$ as
$\xi_\text{B}^2=1+2\left(\langle a^\dagger a \rangle-\langle a^2\rangle\right)$.
Then, a brief summary of this expression is provided below
\begin{equation}\label{xi_approx}
\xi_\text{B}^2=1+\sum_{n=0}^{\infty} K_{n}^{n+2},
\end{equation}
which is proper at low-temperature steady state. 
It is also shown at Eq.~(\ref{xi_approx_appendix})  in Appendix \ref{appendix1}.
$K_{n}^{n+2}$ quantifies the contribution of two-photon process $|n\rangle \leftrightarrow |n+2\rangle$ to the cumulants of $\xi_\text{B}^2$ as
\begin{widetext}
\begin{eqnarray}~\label{xi_K}
K_{n}^{n+2}=  \left[ {n |c_n|^{2}+|c_1|^{2}\delta_{n,1}}+(n+2)|c_{n+2}|^2-2|c_n| |c_{n+2}|\sqrt{(n+1)(n+2)}\right] \nonumber\\
\ =|c_n|^2\left[ (n+2)\left(\frac{|c_{n+2}|}{|c_n|}-\sqrt{\frac{n+1}{n+2}}\right)^2-1+\delta_{n,1}\right],
\end{eqnarray}
\end{widetext}
with $c_{n}$ the probability amplitude in the Fock-state basis. It is obvious that if $K_n^{n+2}<0$ ($K_n^{n+2}>0$), the corresponding process $|n\rangle \leftrightarrow |n+2\rangle$ will contribute to photon squeezing positively (negatively). In particular, due to the robust positive value of $K_{1}^{3}$ unveiled by the Eq.~(\ref{xi_K}), the two-photon process $|1\rangle\leftrightarrow|3\rangle$ tends to destroy photon squeezing.
On the contrary, $K_{0}^{2}$ is almost always negative for $|c_2|/|c_0|< \sqrt{2}$ in most cases, which makes photon squeezing benefiting from $|0\rangle\leftrightarrow|2\rangle$. Thus, the two-photon process $|0\rangle\leftrightarrow|2\rangle$ is the unique contributor to photon squeezing in the low-excitation regime with at most three photons, whereas the process $|1\rangle\leftrightarrow|3\rangle$ destroys photon squeezing. Also, the results of Eq.~(\ref{xi_approx}) based on numerical data of ground-state $|c_n|$ are compared with numerical steady-state $\xi_\text{B}^2$. 

\begin{figure}[tbp]
\centering
\includegraphics[
width=\linewidth]{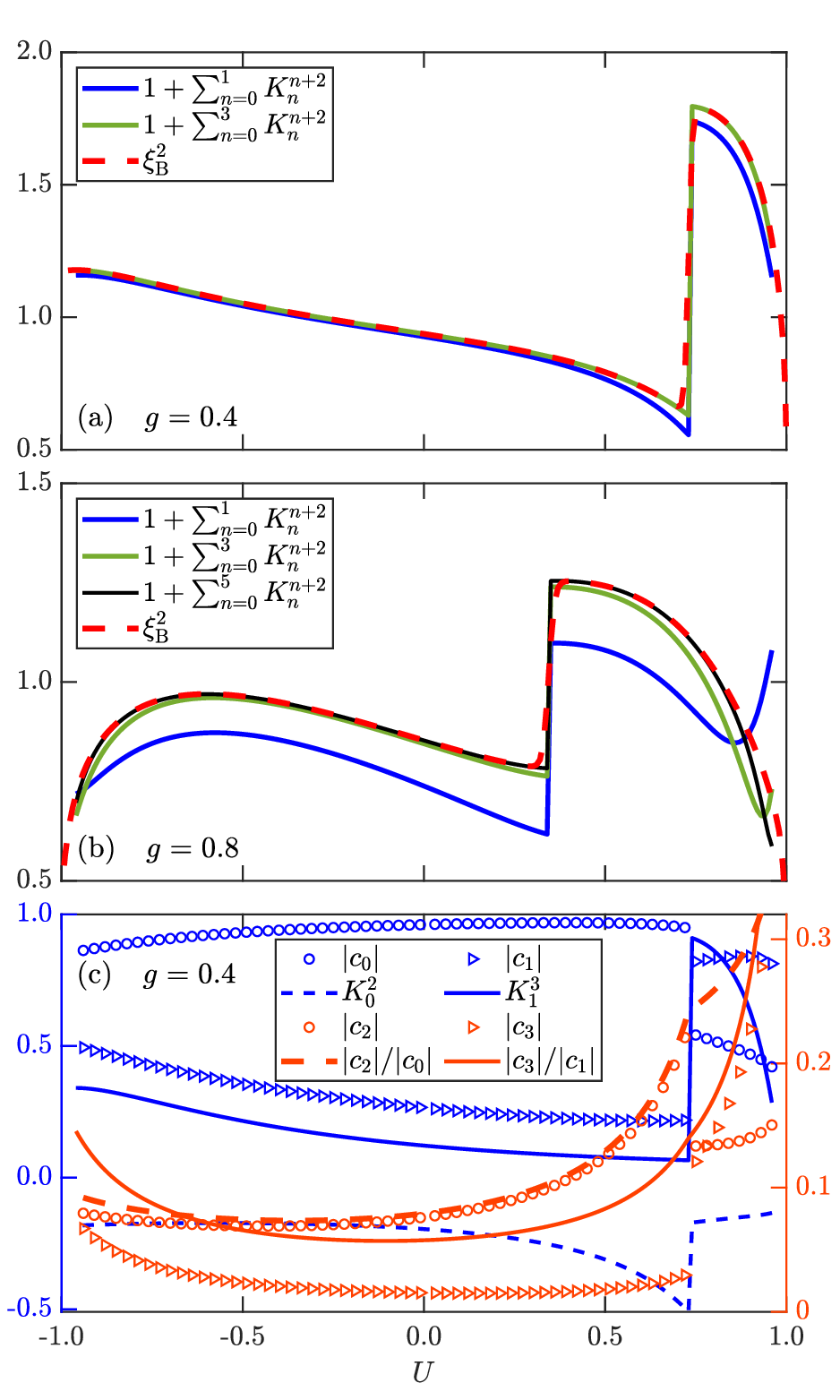}
\caption{(a) and (b)~The comparison between the results based on Eq.~(\ref{xi_approx}) and the numerical one. 
(c)~ground-state probability amplitudes $|c_n|$, its ratios and two major two-photon-process components of $\xi_\text{B}^2$ involved in Eq.~(\ref{xi_approx}); Besides, the data refer to the same-color axis in dual-y-axis plot.  The other parameters not mentioned here are the same as those in Fig.~\ref{fig1}.}
\label{fig2}
\end{figure}

Fig.~\ref{fig2}(a) shows that the blue line with only $K_{0}^2$ and $K_{1}^3$ can describe the variation of numerical $\xi_\text{B}^2$ quite well.
Once higher-excited terms $K_{2}^4$ and $K_{3}^5$ are included, the corresponding green line overlaps perfectly  with the red numerical one. 
If the dipole coupling evoking photonic excitation is strengthened further, as shown in Fig.~\ref{fig2}(b) with $g=0.8$, it is evident that a few more terms of ${K_{n}^{n+2}}$ can match the numerical counterpart.

Notably, the result of $1+K_{0}^2+K_{1}^3$ comparatively underestimates $\xi_\text{B}^2$,
whereas it captures the sharp transition associated with the first-order QPT, implying the dominant  influence of the Stark coupling imposed on the system is not caused through higher-excited terms but primarily by $K_{0}^2$ and $K_{1}^3$. Motivated  by those facts, we will elucidate  the influencing mechanism of the Stark coupling on photon squeezing based on the two-photon-process component $K_{0}^2$ and $K_{1}^3$ in the next paragraph.

Compared with the standard  QRM, a positive Stark coupling makes the ground state (with even parity before the first-order QPT) favor $|-,2\rangle$ but not $|+,1\rangle$, in the subspace with two-total excitation number.
This is because positive Stark term $U\sigma_z a^\dagger a$ results in easier excitation of the polaritons. 
It then implies that enhancing $U$ will increase $|c_2|$ and decrease $|c_1|$, as shown in Fig.~\ref{fig2}(c). 

Next, by inspecting Eq.~(\ref{xi_K}), it is straightforward to find that $|K_n^{n+2}|$ is strengthened with increasing $|c_n|$
(i.e., $K_0^{2}$ decreases for $K_0^{2}<0$ and $K_1^{3}$ increases for $K_1^{3}>0$); 
whereas $K_n^{n+2}$ decreases with increasing $|c_{n+2}|$, for the ratio $|c_{n+2}|/|c_{n}|$ is generally less than $\sqrt{(n+1)/(n+2)}$ in low-excitation regime, 
\emph{e.g.}, $|c_{n+2}|/|c_{n}|\lesssim0.3$ with $g=0.4$ as shown in Fig.~\ref{fig2}(c). 
Thus, the increase of $|c_2|$ induced by positive Stark coupling  strengthens  $|K_{0}^2|$, while, decreasing $|c_1|$  weakens  $K_{1}^3$. Besides, both $|c_0|$ and $|c_3|$ remain nearly unchanged, so the influence can be neglected. 

Therefore, the constructive process $|0\rangle\leftrightarrow|2\rangle$ and the destructive process $|1\rangle\leftrightarrow|3\rangle$ cooperatively lead to stronger squeezing of photons in the positive-$U$ QRSM  compared with the standard Rabi model as shown in upper middle of Fig.~\ref{fig1}(a).

The suppression of photonic squeezing for a negative Stark coupling stems from analogous mechanism.  A negative Stark coupling tends to increase $|c_1|$ comparing with $|c_2|$, since it favors $|+,1\rangle$ but not $|-,2\rangle$, in the two-excitation number subspace.  It then relatively suppresses squeezing by enhancing the destructive process $|1\rangle\leftrightarrow|3\rangle$, as shown by two blue lines in Fig.~\ref{fig2}(c).  Consequently, photon squeezing recedes in the negative-$U$ QRSM, as shown in lower middle panel of Fig.~\ref{fig1}(a).

Finally, we turn to discuss the sharp transition  of the photon field from the squeezing state to the normal state, when the system crosses the first-order QPT, as denoted by the black solid line in Fig.~\ref{fig1}(a). 
The ground-state parity converts from even to odd across the first-order QPT, which has a decisive effect on the sharp transition, which results in the ground state converting from an even-excitation number to an odd one, i.e., from the subspace spanned by $\{\vert-,2n\rangle,\vert+,2n+1\rangle\}$ to that spanned by  $\{\vert+,2n\rangle,\vert-,2n+1\rangle\}$.
Therefore, the squeezing contributor switches from $|-,0\rangle\leftrightarrow|-,2\rangle$ to $|+,0\rangle\leftrightarrow|+,2\rangle$, and  the destructive process shifts from $|+,1\rangle\leftrightarrow|+,3\rangle$ to $|-,1\rangle\leftrightarrow|-,3\rangle$. 
As this occurs, the populations $|c_0|$ and $|c_2|$ decrease ($|c_1|$ and $|c_3|$ increase) dramatically for the opposite changing trend of the total-excitation number of the corresponding basis states. 
Thus, from Eq.~(\ref{xi_K}), the downward (upward) jumps between $|c_0|$ and $|c_1|$ lead to sharp suppression (enhancement) of the squeezing constructive process $|0\rangle\leftrightarrow|2\rangle$  (destructive process $|1\rangle\leftrightarrow|3\rangle$ ). This demonstrates that the vanishing of photon squeezing across the first-order QPT originates from the parity switching in the ground states.

\section{\label{sec4}Strong photon squeezing induced by the superradiant phase transition}

We turn to analyze the influence of the second-order SRPT on photon squeezing.
Fig.~1(b) already shows that the quadrature squeezing factor $\xi_\text{B}^2$
decreases to a minimum as the linear coupling strength $g$ approaches the critical points  $g_c^{+}=0.5$ and $g_c^{-}=\sqrt{3/4}\approx0.86$    for $U$ near $\pm 1$, respectively. The critical points are marked and labeled on the horizontal axis in Fig.~\ref{fig1}(c)~\cite{yfxie2019jpa,yfxie2020pra,xychen2021pra}.
The effective size is defined as $L={1}/{(1-|U|)}$~\cite{yfxie2020pra,xychen2021pra}.

Those facts suggest the possible critical behavior of the quadrature squeezing factor $\xi_\text{B}^2$ in spite of 
the fact that $\xi_\text{B}^2$ does not correspond to a single physical observable. It is not surprising for the fact 
that the SRPT implies the macroscopic excitation in a certain field quadrature with large fluctuation $\langle X_{\theta_{0}}^2\rangle$ ~\cite{hwang2015prl,xychen2021pra,mxliu2017prl,tianye2025pra}, while $\langle X_\theta\rangle=0$ with arbitrary $\theta$ in the finite size system for the $\mathcal{Z}_2$ symmetry. 
Then, the nearly vanishing variance of the field quadrature is confined to the canonically conjugate one, namely, $\xi_\text{B}^2=\langle P_{\theta_{0}}^2\rangle$. Indeed, such criticality-induced photon squeezing has also been mentioned in the standard QRM and Dicke model~\cite{squeeze_critical_Dicke,hwang2018pra,xdh2026pra,squeeze_rabi}.

In contrast to the above earlier scenarios,  the strong squeezing of photons observed here is almost robust even in the deep superradiant phase, which should facilitate the preparation of strongly squeezing states in the experiment.

\begin{figure}[tbp]
\centering
\includegraphics[width=\linewidth]{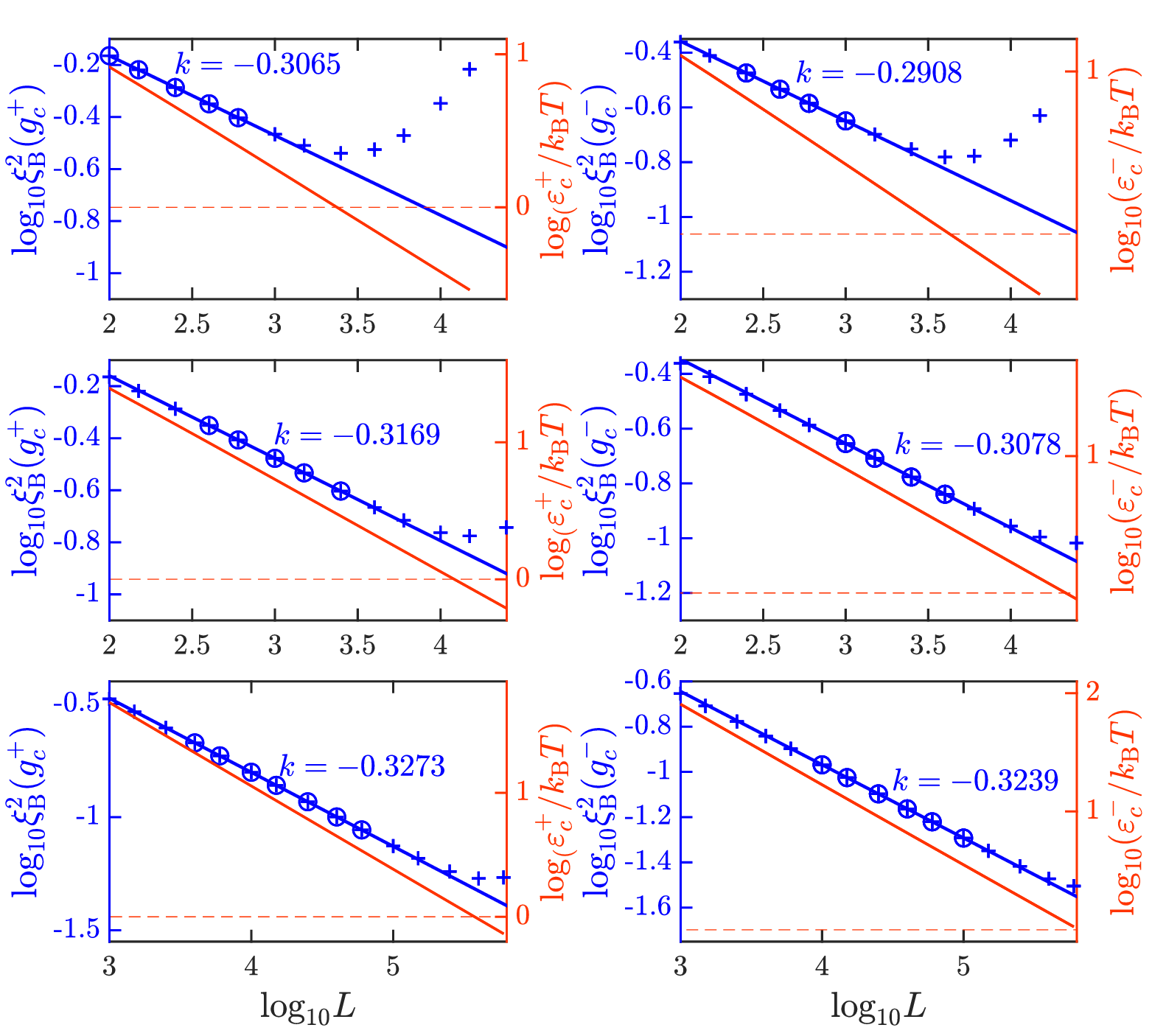}
\caption{Power-law-type scaling behaviors of the squeezing factor $\xi_\text{B}^2\propto L^{-\gamma}$  and its crossover out of the quantum critical regime driven by thermal fluctuation. In the dual-y-axis plot, the numerical data of $\xi_\text{B}^2(g_c^\pm)$ are shown with  blue plus signs on the right axis, while the ratio $\varepsilon_c^{\pm}/k_\text{B}T$  with $\varepsilon_c^{\pm}$ the energy gap of the two lowest energy level is shown on the left axis. Moreover, the linear fitting $\xi_\text{B}^2(g^{\pm}_c)\propto L^{k}$ is shown with the blue solid lines, accompanied by the fitted data emphasized by same-color circle. Besides, from top to bottom, the temperatures of two bathes decreases, i.e., $k_\text{B}T=0.003\omega_0$, $0.001\omega_0$ and $0.0001\omega_0$ with $T_\text{q}=T_\text{c}=T$ for the first, second, and third row respectively. The other parameters  are the same as those in Fig.~\ref{fig1}.}
\label{fig3}
\end{figure}

We  investigate the critical behavior of the quadrature squeezing factor $\xi_\text{B}^2$ 
through the  power-law-type scaling law 
\begin{equation}\label{xi_c_scaling}
\xi_\text{B}^2\left(g_c^{\pm}\right)\propto L^{-\gamma}
\end{equation}
where $\xi_\text{B}^2\left(g_c^{\pm}\right)$ is the value at the critical coupling of the second-order QPT, with $L={1}/({1\mp U})$ the effective size and $\gamma$ the scaling exponent. 

According to the scaling law (\ref{xi_c_scaling}), Fig.~\ref{fig3} is plotted with decreasing environment temperatures from top to bottom.
The numerical results of $\xi_\text{B}^2\left(g_c^{\pm}\right)$ for different values of $L$ are denoted by  blue plus signs on the left axis in the dual-y-axis of Fig.~\ref{fig3}. The solid blue line represents the linear fitting,  with the fitted data are emphasized by same-color circle and the fitted slopes $k$, defined by $\xi_\text{B}^2\left(g_c^{\pm}\right){\propto}L^{k}$, are labeled in the same color. The ranges of those linear fittings are chosen to satisfy the residual norms less than $5\times 10^{-4}$.

It is then clear that the scaling law~(\ref{xi_c_scaling}) holds with the scaling exponent $-\gamma$ approximated as the fitted slope $k$ for a wider range of the effective size as the environment temperature decreases. 
This indicates that the steady state of the open QRSM may present arbitrarily strong squeezing at sufficiently low environment temperature and sufficiently large effective size,  providing a promising route to realizing strong optical squeezing states scalable with the effective size in experiments. 

Furthermore, the fitted slopes $k$  show a tendency $k\rightarrow-1/3$ as  environment temperature decreases, implying the ground-state critical exponent $\beta=\gamma/\nu=1$, assuming  the correlation-length exponent $\nu=1/3$ as in ground-state counterpart  and with $\gamma\approx -k$~\cite{xychen2021pra}. Obviously, it is consistent with the analytical result of $U=1$ in Appendix \ref{appendix2}. Overall, we have shown the
quantum criticality of the photon-squeezing factor $\xi_\text{B}^2$ through the numerical power-law behavior.

We now turn to the crossover driven by thermal fluctuation. 
The crossover may be characterized by the upper limits of the above size ranges, where the scaling law~(\ref{xi_c_scaling}) holds. 
To show this thermal-fluctuation-driving crossover clearly, the ratio of two key energy scales, i.e., $k_\text{B}T$ for thermal fluctuation and the energy gap $\epsilon_c^{\pm}$ of the two lowest energy levels at the critical point of the SRPT $g_c^{\pm}$, is illustrated with the orange solid line on the right axis in the dual-y-axis of Fig.~\ref{fig3}. Evidently, $\epsilon_c^{\pm}$ vanishes following the similar power law of the energy gap $\epsilon_c^{\pm}\propto L^{-2/3}$~\cite{xychen2021pra}. 
This implies that there always exists a sufficiently large system size $L$, such that $\epsilon_c^{\pm}=k_\text{B}T$  at the cross point of two orange lines. 

Then, it is easy to find the tendency that $\xi_\text{B}^2\left(g_c^{\pm}\right)$, denoted by the plus sign, begins to deviate from the solid blue fitted line,
as $\epsilon_c^{\pm}$ becomes close to $k_\text{B}T$.
Once the condition $\epsilon_c^\pm\lesssim k_\text{B}T$ is satisfied, $\xi_\text{B}^2\left(g_c^{\pm}\right)$ departs entirely from the solid blue fitted line. 
Considering the fact that the fitted line represents the power-law-type scaling law~(\ref{xi_c_scaling}), the above phenomena imply that the condition $\epsilon_c\approx k_\text{B}T$ serves as a suitable upper limit of the size range for the scaling law, and further for the quantum critical regime in the presence of thermal fluctuations. 

\begin{figure}[tbp]
\centering
\includegraphics[width=\linewidth]{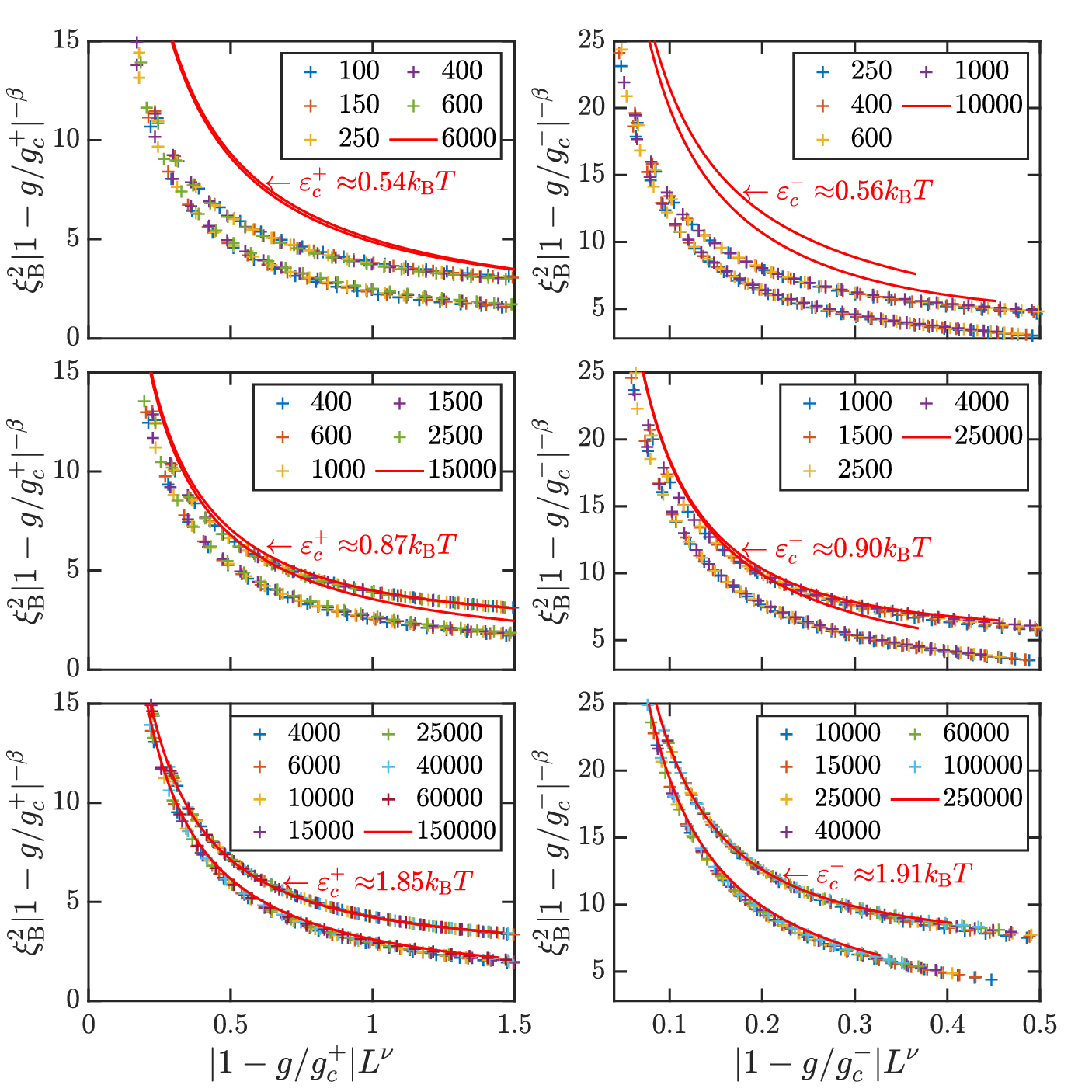}
\caption{ Finite-size scaling of the squeezing factor $\xi_\text{B}^2$ and its crossover out of the quantum critical regime driven by thermal fluctuation. The numerical data of all effective size $L$ fitted linearly in Fig.~\protect \ref{fig3} \protect are denoted with different-color plus signs, showing the finite-size scaling behaviors consistent with expectations, while the red lines shows the evident deviating from the scaling function \protect(\ref{xi_FSS}) \protect along with the near-$k_\text{B}T$ value of $\epsilon_c^{\pm}$ texted in the same color. In the legend, we shown the effective sizes of different curves. Besides, the environment-temperature distribution across rows is same as that in Fig.~\protect\ref{fig3}\protect, i.e., $k_\text{B}T/\omega_{0}=0.003$, $0.001$ and $0.0001$ with $T_\text{q}=T_\text{c}=T$ for the first, second, and third row respectively. The other parameters not mentioned here are the same as those in Fig.~\protect\ref{fig1}\protect.}
\label{fig4}
\end{figure}

To confirm quantum criticality and its crossover under thermal fluctuation further, we resort to the finite-size-scaling function in the critical regime of continuous QPT. The finite-size-scaling function of $\xi_\text{B}^2$ may be assumed as
\begin{equation}\label{xi_FSS}
\xi_\text{B}^2=\vert 1-g/g_c^{\pm}\vert^\beta f\left(\vert 1-g/g_c^{\pm}\vert L^{\nu}\right),
\end{equation}
where $\nu$ is critical exponent of correlation length assumed as ground-state counterpart $\nu = 1/3$~\cite{xychen2021pra}, and $\beta$ is the critical exponent of $\xi_\text{B}^2$ satisfying the scaling law of exponents $\beta=-\gamma/\nu$.

Based on the scaling function (\ref{xi_FSS}), Fig.~\ref{fig4} is plotted with $-\gamma$ approximating as the slope $k$ of the linear fitting in Fig.~\ref{fig3}. The numerical results denoted with different-color plus sign keep the same size as those exhibiting power-law behavior and linear fitted in Fig.~\ref{fig3}. 
It is clear that all plus-sign curves for these sizes collapse well onto a single one, confirming the validity of the assumed  finite-size-scaling function (\ref{xi_FSS}) of $\xi_\text{B}^2$. 

Moreover, a red curve corresponding to the size for which  $\varepsilon_c^\pm\approx k_\text{B}T$ is shown in each subfigure to illustrate the deviation from the finite-size scaling function (\ref{xi_FSS}). 
In particular, as  $\varepsilon_c^\pm/k_\text{B}T$ increases from top to bottom in Fig.~\ref{fig4}, the red curves approach the plus-sign curve governed by the finite-size function \ref{xi_FSS} gradually. It confirms the condition $\varepsilon_c\approx k_\text{B}T$ setting a reasonable upper size limit for the quantum critical regime of the finite-size system. This condition may also serve as a practical reference of the environment temperature when preparing strongly squeezing states in the experiment.

\section{\label{conclusion} Conclusion}
In the paper, we investigate steady-state optical squeezing in the  open QRSM under thermal noise in the framework of the quantum dressed master equation, which enables us to
treat the arbitrarily strong qubit-photon coupling reasonably. The quadrature squeezing of the steady-state photon field is then numerically calculated, as shown in Fig.~\ref{fig1}. The results reveal clear trends that a positive Stark coupling tends to enhance the squeezing, whereas a negative Stark coupling tends to suppress it over the main parameter region.

In particular, the quadrature squeezing exhibits a characteristic signal associated with both first- and second-order QPTs. 
The optical squeezing becomes sharp vanishing once the first-order QPT occurs,  whereas the squeezing becomes pronounced and robust in the critical regime of the second-order QPT near $U\approx \pm 1$.

Then, we derive an analytical expression of the quadrature squeezing, based on the intrinsic property of the ground state. This expression splits the expectation value of the quadrature squeezing into a series of components that are tightly related to the two-photon processes $\{ \vert n\rangle\leftrightarrow \vert n+2\rangle\}$ in the Fock-state basis. The dominant components at low temperature exhibit opposite effects, i.e.,  $ \vert 0\rangle\leftrightarrow \vert 2\rangle$
positively contributes to the squeezing, while  $ \vert 1\rangle\leftrightarrow \vert 3\rangle$ suppresses the squeezing. 
We also unveil the mechanism underlying the almost opposite influences of positive and negative Stark couplings, and the sharp vanishing of the squeezing across the first-order QPT. 

Finally, we analyze the significantly enhanced photon squeezing associated with the second-order SRPT. We first show the criticality-enhanced squeezing, which is scalable with the effective system size through  power-law-type scaling behaviors at the critical point. We demonstrate that the collapse of curves for different effective system sizes onto a single universal curve is governed by the finite-size-scaling function in the critical regime.
Furthermore, we find that the destruction of quantum criticality by thermal fluctuation can be quantified by the condition $k_\mathrm{B}T>\varepsilon_c$, where $\varepsilon_c$ is the energy gap at the critical point, which may serve as a practical  reference for experiments. 
These results may be helpful to deepen the understanding of nonclassical light in qubit-photon coupled systems. 
In the future, our proposal may also provide a promising platform for preparing the optical squeezing state.

\section*{Acknowledgements}
This work was supported by the Zhejiang Provincial Natural Science Foundation of China under Grant No. LZ25A050001.

\appendix
\section{An analytical expression of quadrature squeezing for the ground state of the QRSM\label{appendix1}}
In this section of the appendix, we derive an analytical expression of $\xi_\text{B}^2$. For this purpose, the minimization in the quadrature-squeezing-factor definition (\ref{xi_def}) should be performed. A general minimization provides the expression~\cite{jianma2011physrep}
\begin{equation}\label{xi_min1}
\xi_\text{B}^2=1+2\left(\langle a^\dagger a\rangle-\vert\langle a\rangle\vert^2\right)-2\vert\langle a^2\rangle-\langle a\rangle^2\vert,
\end{equation}
which is indeed difficult to process analytically for the modulus of expectations.
We thus explore steady-state properties to simplify the photon squeezing expression. Due to the quite low temperature of the environment, the steady state is almost the ground state of the QRSM. The ground state can be expressed as
\begin{equation}\label{phi0_full}
|\phi_0^{\pm}\rangle= \sum_{n=0}^{\infty} [{c_{2n} \vert\pm,2n\rangle+c_{2n+1} \vert\mp,2n+1\rangle}],
\end{equation}
where $\vert n\rangle$ denotes the Fock state, $\vert+\rangle$ ($\vert-\rangle$) denotes the higher (lower) eigenstate of the qubit and all probability amplitudes $c_n$ may be taken as real values. It need to note that the parity symmetry  has been considered, i.e., the even (odd) parity denoted by the superscript $+\ (-)$ of $\phi$ confines the  ground state populating in the subspace with even (odd) total excited number. 

By inspection of the ground-state expansion~(\ref{phi0_full}), it is easy to obtain $\langle a\rangle=0$. 
Then, to determine the sign of $\langle a^2\rangle$, we calculated the expectation of the Hamiltonian (\ref{HRS}) with the ground-state expression (\ref{phi0_full}),
\begin{eqnarray}
E_0^{\pm}& =\sum_{n=0}^{\infty} \left[H_n^{\text{D}} +H_n^{g}\right].
\end{eqnarray}
In the expression, $H_n^{g}$ is the cross term produced by the dipole-coupling term, while the $H_n^\text{D}$ is the diagonal term, i.e., 
\begin{subequations}\label{E0}  
\begin{align}
H_n^{\text{D}}&=c_{2n}^2 \left[2n(\omega\pm U)\pm\frac{\Delta}{2}\right] \notag \\
&+c_{2n+1}^2 \left[(2n+1)(\omega\mp U)\mp\frac{\Delta}{2}\right], \\
H_n^{g}&=2\sqrt{2n+1}gc_{2n}c_{2n+1}+\sqrt{2n}gc_{2n}c_{2n-1} \notag \\
&+\sqrt{2n+2}gc_{2n+1}c_{2n+2}.
\end{align}
\end{subequations}
Thus, the ground-state probabilities $c_n$ and $c_{n+1}$ should process opposite signs, and ${c_{n}}$ and ${c_{n+2}}$ same sign for the fact that the ground state should minimize the Hamiltonian expectation.
It further implies that the expectation $\langle a^2\rangle$ is non-negative. 
Then,  we simplify the general expression~(\ref{xi_min1}) of $\xi_\text{B}^2$ as
\begin{equation}
\xi_\text{B}^2=1+2\left(\langle a^\dagger a\rangle-\langle a^2\rangle\right). \label{xi_smplf}
\end{equation}

Next, considering the fact that $\langle a^2\rangle$ is associated with two-photon processes, we divide both  $\langle a^\dagger a\rangle$ and  $\langle a^2\rangle$ into a series of components associated with the two-photon processes $\{|n\rangle \leftrightarrow |n+2\rangle\}$, which provides the expression of $\xi_\text{B}^2$ as follows,
\begin{equation}\label{xi_approx_appendix}
\xi_\text{B}^2=1+\sum_{n=0}^{\infty} K_{n}^{n+2},
\end{equation}
where $K_{n}^{n+2}=\mathrm{Tr}\{(a^\dagger a-2a^2)P_{n,n+2}|\phi_0\rangle\langle\phi_0|\}+\delta_{n,1}|c_1|^2$, with $P_{n,n+2}$ the projector operator of the subspace spanned by the Fock states $|n\rangle$ and $|n+2\rangle$. The vanishing $2$ before $a^\dagger a$ and the modification $\delta_{n,1}|c_1|^2$ are attributed to the fact that all $\langle a^\dagger a\rangle$'s components in n-photon subspace are counted repetitively in both $K_{n-2}^n$ and $K_n^{n+2}$ except the one-photon-subspace component merely.
Then, the contribution of two-photon process $|n\rangle \leftrightarrow |n+2\rangle$ to the cumulants of $\xi_\text{B}^2$ is quantified as
\begin{widetext}
\begin{subequations}\label{xi_K_appendix}  
\begin{align}
K_{n}^{n+2}=  \left[ {n |c_n|^{2}+|c_1|^{2}\delta_{n,1}}+(n+2)|c_{n+2}|^2-2|c_n| |c_{n+2}|\sqrt{(n+1)(n+2)}\right] \\
\ =|c_n|^2\left[ (n+2)\left(\frac{|c_{n+2}|}{|c_n|}-\sqrt{\frac{n+1}{n+2}}\right)^2-1+\delta_{n,1}\right],
\end{align}
\end{subequations}
\end{widetext}
where all probability amplitudes are taken as the absolute value for clarity, and it is feasible because $c_n$ and $c_{n+2}$ always process the same sign. Considering the similarity of the low-temperature state and the ground state, it is feasible to analyze the steady-state quadrature squeezing approximately with the above expression.

\section{Quadrature squeezing of photons of the QRSM at $U =1$ \label{appendix2}}
In the appendix, we derive the critical behavior of the quadrature photon squeezing based on the analytical exact solution of the QRSM at $U =1$~\cite{yfxie2019ctp,xychen2021pra}, and the results of $U=-1$ can be obtained from replacing $\Delta$ by $-\Delta$ in those of $U=1$. 

In the basis of the eigenstates $|+\rangle=(1,0)^T$ and $|-\rangle=(0,1)^T$ of $\sigma_z$, the Hamiltonian~(\ref{HRS}) with $U=1$ ($\omega_{0}$ has been taken as unit) can be rewritten in matrix form as
\begin{eqnarray}
H_\text{RS}=
\left(
  \begin{array}{cc}
    2 a^\dagger a+\frac{\Delta}{2} & g\left(a^\dagger+a\right) \\
    g\left(a^\dagger+a\right) & -\frac{\Delta}{2}\\
  \end{array}
\right).
\end{eqnarray}
Similarly, the wave function can be expressed as 
\begin{equation}
|\Psi\rangle= 
\left(
\begin{array}{c}
                |\Psi_{+} \rangle\\
                |\Psi_{-} \rangle
              \end{array}
\right).
\end{equation}
where $\Psi_{\pm}$ denotes the photonic wave function associated with the higher (lower) level of the qubit. Then the Schr\"{o}dinger equation gives
\begin{subequations}\label{sch_eq}
\begin{align}
\left(2a^\dagger a+\frac{\Delta}{2}\right)|\Psi_{+} \rangle+ g(a^\dagger+a) |\Psi_{-} \rangle 
=E|\Psi_{+} \rangle,\\
g(a^\dagger+a)|\Psi_{+} \rangle=\left(\frac{\Delta}{2}+E\right) |\Psi_{-} \rangle. 
\end{align}
\end{subequations}
Substituting Eq.~(\ref{sch_eq}b) into Eq.~(\ref{sch_eq}a), we can remove $|\Psi_{-} \rangle$ and obtain the effective Hamiltonian for $|\Psi_{+} \rangle$,
\begin{equation}\label{Heff}
H_\text{eff}=2a^\dagger a+\chi (a^\dagger +a)^2+\frac{\Delta}{2}
\end{equation}
with $\chi=g^2/(\frac{\Delta}{2}+E)$. Next, resorting to the squeezing transformation $S = e^{r\left(a^2-{a^\dagger}^2\right)/2}$ with $r=\frac{1}{4}\ln \left( \frac{1}{1+2\chi}\right)$, we can diagonalize the effective Hamiltonian~(\ref{Heff}) and get a quantum-oscillator-type one
\begin{equation}\label{Hdiag}
H^{'}=SH_\text{eff}S^\dagger=\sqrt{1+2\chi}\left(2a^\dagger a+1\right)-1+\frac{\Delta}{2}.
\end{equation}
Thus, the eigenenergy is obtained as
\begin{equation}\label{EU1}
E_n=\sqrt{1+2\chi_n}(2n+1)-1+\frac{\Delta}{2}.
\end{equation}
with $n=0,1,2,\dots,\infty$ for different eigen-levels from lowest energy to highest energy, 
and the eigenfunction in the Fock basis $\{|n\rangle\}$ reads
\begin{equation}\label{phin}
|\phi_{n}\rangle= 
\frac{1}{N_n}\left(
\begin{array}{c}
                e_n S^\dagger|n\rangle\\
                d_n S^\dagger\left(a^\dagger +a\right)|n\rangle
\end{array}
\right).
\end{equation}
with the coefficients $e_n=(1+2\chi_n)^{1/4}$, $d_n=\chi_n/g$, and the normalization factor $N_n=\sqrt{c_n^2+(2n+1)d_n^2})$.

Equation~(\ref{EU1}), which provides the eigenenergy, is a nonlinear equation for the $E$-dependence of $\chi_n=g^2/(E_n+\frac{\Delta}{2})$, which has no general analytical solution. In the critical regime, all low-energy levels tend to collapse at $E_c^{+}=-\Delta/2-2g^2$ when the critical point $g_c^{+}=\sqrt{(1-\Delta)/2}$ is approached ~\cite{xychen2021pra,yfxie2019jpa,yfxie2019ctp}. This causes $\chi_n$ to decrease to $-\frac{1}{2}$, making $\sqrt{1+2\chi_n}$ an infinitesimal quantity. Given this, Eq.~(\ref{EU1}) can be solved by fixed-point iterating the rewritten equation \begin{equation}
z=f(z)=\frac{1}{2n+1}\left[2g^2/(z^2-1)+2{g_c^{+}}^2\right]
\end{equation} 
with $z=\sqrt{1+2\chi_n}$, in the neighborhood of $z=0$, where global convergence is guaranteed by $f^{\prime}(z)\propto z<1$. Taking the initial value for the fixed-point iteration as $z_0=0$, $z_1$ and $z_2$ is given as
\begin{equation}\label{z_eq}
z_{1}=\frac{2\delta}{2n+1},z_{2}=z_1-\frac{2g^2}{2n+1}z_1^2+O(\delta^4),
\end{equation}
where $\delta={g_c^+}^2-g^2$ is a small quantity in the critical regime. The second iteration provides $z_2$ just with a higher-order infinitesimal correction $z_2-z_1\propto z_1^2\propto\delta^2$, implying the convergence  of the fixed-point iteration. Thus, $\sqrt{1+2\chi_n}$ and $\chi_n$ 
involved in the eigenfunction (\ref{phin}) are given as
\begin{eqnarray}\label{chi_eq}
\sqrt{1+2\chi_{n}}=\frac{2\delta}{2n+1}-\frac{8g^2\delta^2}{(2n+1)^3}+\mathcal{O}(\delta^3) \\
\chi_{n}=-\frac{1}{2}+\frac{2\delta^2}{(2n+1)^2}+\mathcal{O}(\delta^4).
\end{eqnarray}
Then the coefficients of the wavefunction are
\begin{eqnarray}\label{psin_en}
\frac{e_n}{N_n}=\frac{2g\sqrt{2\delta}}{(2n+1)}\left[1-\frac{4g^2\delta}{(2n+1)^2}+\mathcal{O}(\delta^3)\right], \\
\frac{d_n}{N_n}=-\frac{1}{\sqrt{2n+1}}\left[1-\frac{4g^2\delta}{(2n+1)^2}+\mathcal{O}(\delta^2)\right].
\end{eqnarray}

Finally, the variance of the momentum quadrature in the eigenstate is calculated as 
\begin{eqnarray}\label{xi_U1}
(\Delta P)^2 &=& \left(2n+1\right)e^{-2r}\frac{e_n^2}{N_n^2}+\left(2n^2+2n+3\right)e^{-2r}\frac{d_n^2}{N_n^2} \notag\\
&=&\frac{2\delta(2n^2+2n+3)}{(2n+1)^2}+\mathcal{O}(\delta^2).
\end{eqnarray}
Thus, $(\Delta P)^2$ vanishes proportionally to $ (g_c^+-g)$ near the critical point, since $\delta={g_c^+}^2-g^2$. This clearly shows the criticality of ground-state $\xi_\text{B}^2$, which is minimized in the direction of momentum quadrature with the exponent $\beta=1$.

\end{document}